\documentstyle[12pt]{article}
\begin{document}
\baselineskip 10mm


\vskip 4mm

\centerline{\large \bf Metastable spin-polarized carbon clusters C$_8$
and their ensembles}

\vskip 2mm

\centerline{A. A. Izmalkov, L. A. Openov}

\vskip 2mm

\centerline{\it Moscow Engineering Physics Institute
(State University),}
\centerline{\it Kashirskoe sh. 31, Moscow 115409, Russia}

\vskip 4mm

\begin{quotation}

Results of tight-binding calculations on metastable spin-polarized carbon
clusters C$_8$ (prismanes) are presented. It is shown that those clusters can
form ensembles due to intercluster bonding. The binding energy of a given
metastable configuration decreases monotonously with the total spin $S$,
while the activation energy for the decay of a metastable state has distinct
maxima as a function of $S$. For specific values of $S$, the
dynamical stability of spin-polarized ensembles of prismanes appears to be
higher than that of an isolated prismane, pointing to a possibility of
existence of "cluster matter" composed of spin-polarized clusters C$_8$.

\end{quotation}

\vskip 6mm

{\bf 1. Introduction}

\vskip 2mm

Carbon is a chemical element with a great number of unique
properties. In particular, carbon is known by its polymorphism. Due
to its ability to exist in different valence states, carbon can
form a rich variety of compounds, including diamond, graphite,
carbyne, etc. In the field of nanostructures, cage-like fullerenes
(C$_{60}$ etc. \cite{Kroto}) attract much attention due to their
both practical and fundamental importance. Recently, the cage-like
cluster C$_{20}$ was synthesized \cite{Prinzbach}. This cluster is
the smallest possible fullerenes cage consisting of 12 pentagons.
First-principle calculations \cite{Okada} point to a possibility of
synthesis of condensed C$_{20}$ phases with three-dimensional
covalent networks.

In recent numerical
simulations of small carbon clusters, a possible existence of another
cage-like cluster C$_{8}$ (prismane) was suggested theoretically
\cite{Openov}. Prismane was predicted to have the shape of a six-atom
triangular prism with two excess atoms above and below its bases, see Fig.1.
There is a good reason to believe that prismane C$_8$ is the smallest
possible {\it three-dimensional} carbon cluster.
Prismane is the {\it metastable} cluster. Its binding energy $E_b$ (defined
as $E_{b}(N)=NE(1)-E(N)$, where $E(N)$ is the total energy of an
$N$-atom cluster, $E(1)$ is the energy of an isolated carbon atom) was
calculated to be 5.1 eV/atom \cite{Openov}, i.e., 0.45 eV/atom lower than the
binding energy of the stable one-dimensional eight-atom cluster
and 2.3 eV/atom lower
than the binding energy of the bulk graphite or diamond. It is essential,
however, that the calculated value of the lowest energy barrier that prevents
the prismane from decay into the stable or another metastable atomic
configurations with greater values of $E_b$ appears to be rather high, about
$0.44$ eV \cite{lowdim}, resulting in a rather long lifetime of the prismane
even at room temperature. Hence, it is likely that this cluster may be
observed experimentally.

Since the prismane had been predicted \cite{Openov}, the bonding between
prismanes was studied and several configurations in which prismanes form
metastable ensembles were found \cite{Molecules,IWFAC,PLDS}. In those
ensembles, the inter-prismane
bonds are rather strong, so that one could expect prismanes C$_8$ to form a
covalent solid. However, the stability (and hence, the lifetime) of the
prismane-based ensembles was shown to decrease strongly with the number of
prismanes in the ensemble.

Until now, theoretical studies of prismanes and their ensembles were
restricted to the non-polarized case, i.e., the total spin $S$ of the systems
under consideration was assumed to be zero
\cite{Openov,lowdim,Molecules,IWFAC,PLDS}. Meanwhile, the polarization of
small species of substance (including molecules and clusters) can increase
the stability of the system and/or result in the formation of new metastable
configurations. For example, two helium atoms, being in the ground state with
$S=0$ each, do not form the covalently bound molecule He$_2$, while the
existence of metastable long-lived molecule He$_2^*$ in the excited state
with $S=1$ has been demonstrated both theoretically \cite{Yarkony}
and experimentally \cite{McKinsey}. In general, one can expect that the
binding energy of the system composed of nonmagnetic atoms decreases with
$S$, and hence spin-polarized clusters and their ensembles can be considered
as candidates to the so called high energy density materials (HEDM). The
question is whether the lifetimes of such systems are long enough for they
could be synthesized, investigated and used in practice. So, the study of
metastable spin-polarized atomic configurations is of great fundamental
interest and practical importance.

In this paper, we extend our previous studies to the case of spin-polarized
prismanes C$_8$ and their ensembles. We show that metastable prismane-like
carbon clusters with $S\neq 0$ can exist and form ensembles due to
intercluster bonding. The binding energies and the activation energies for
the decay of metastable states are calculated as functions of $S$ for an
isolated prismane and different arrangements of prismanes in the ensembles.

\vskip 6mm

{\bf 2. Computational details}

\vskip 2mm

The search for metastable configurations of prismane-like carbon clusters and
their ensembles with different spins $S$ was carried out by means of
structural relaxation technique. Starting with a particular atomic
arrangement, all atoms were allowed to displace step by step in the directions
of interatomic forces, atomic velocities being terminated at each step. The
forces were determined via calculation of the total energy of the system (or,
equivalently, the binding energy) making use of a transferable tight-binding
potential \cite{Xu1,Xu2} that had been proven to reproduce accurately the
energy-versus-volume diagram of carbon polytypes and to give a good
description of both small clusters and bulk structures of carbon
\cite{Xu1,Xu2}. This potential has been used for simulations of metastable
prismane-like structures with $S=0$ in our previous works
\cite{Openov,lowdim,Molecules,IWFAC,PLDS}. The numerical technique employed
is not as sophisticated and time-consuming as {\it ab initio} ones and allows
for molecular dynamics simulations of rather large systems for a relatively
long time.

Depending on the choice of initial configuration, the resulting atomic
arrangement corresponds to either global or local minimum of
the total energy as a function of atomic coordinates for a given number of
atoms in the system (stable and metastable states, respectively). We
were interested in {\it metastable states} separated from the stable and/or
other low-lying metastable configurations by the energy barriers. An
important characteristic of a metastable configuration is the activation
energy $E_a$ for its decay. For a given temperature $T$, the value of $E_a$
determines the probability of the system decay in a unit of time,
\begin{equation}
W\propto \exp \left( -E_a/k_B T\right),
\label{probability}
\end{equation}
and hence the lifetime $\tau$ of a particular metastable state is
\begin{equation}
\tau\propto\tau_0\exp\left( E_a/k_B T\right),
\label{tau}
\end{equation}
where $k_B$ is the Boltzmann constant, $\tau_0$ is a characteristic time of
the order of an oscillation period of the system ($\sim 10^{-13}$ s for the
C$_8$ clusters and their small ensembles).

The lifetimes $\tau$ at different temperatures were calculated directly by
molecular dynamics simulation of thermally isolated systems. Since the
cluster decay is probabilistic in nature, we accumulated a great statistics
($30\div 100$ trials) in order to get a reliable estimate of the value of
$E_a$ through fitting the results obtained to Eq. (\ref{tau}).

\vskip 6mm

{\bf 3. Results and discussion}

\vskip 2mm

To find three-dimensional spin-polarized prismane-like configurations of
eight carbon atoms, we made use of the non-polarized prismane structure
\cite{Openov} as an initial atomic arrangement in the structural relaxation
procedure. We have found three-dimensional metastable C$_8$ configurations
for $S=1\div 7$, while for $S>7$ all relaxed structures appeared to be
either one- or two-dimensional. Metastable
configurations can be subdivided into "nondistorted" ($S=0,4,7$) and
"distorted" ($S=1,2,3,5,6$) ones. The structure of nondistorted prismanes is
characterized by three different bond lengths $a$, $b$, $c$ (symmetrical
configuration), see Fig.1. In distorted prismanes, the bond lengths are
slightly scattered around those three characteristic lengths, see Fig.2.
From Fig.2 one can see that the values of $a$, $b$, and $c$ (or their average
values for distorted configurations) increase with $S$, so that the cluster
volume increases with $S$.

The binding energy $E_b$ of spin-polarized prismanes strongly decreases with
$S$ from $E_b=5.1$ eV/atom at $S=0$ down to $E_b=2.1$ eV/atom at $S=7$, see
the lower curve in Fig.3. Hence, the energy $E_{acc}$ accumulated in
metastable prismanes increases considerably and exceeds 5 eV/atom at $S=7$
(here $E_{acc}=E_b^{graphite}-E_b$, where $E_b^{graphite}=7.4$ eV/atom is the
binding energy of graphite or diamond, the most stable carbon structures
that are characterized by the maximum value of $E_b$ per atom). Such an
energy could be released upon fusion of a large number of prismanes into the
bulk specimen with the graphite structure, accompanying by the
spin-depolarization transition (i.e., the reduction of the total spin
down to $S=0$).

Fig.4 shows the dependence of activation energy $E_a$ for the decay of
metastable spin-polarized prismane C$_8$ versus the total spin $S$. One can
see that the value of $E_a$ is maximum, $E_a=(0.84\pm 0.05)$ eV, for the
non-polarized prismane C$_8$ with $S=0$. An estimate of the lifetime $\tau$
making use of Eq. (\ref{tau}) gives $\tau \sim 1$ s at $T=273$ K (see also
Ref. \cite{Openov}), the value of $\tau$ being increased exponentially with
decreasing $T$. For spin-polarized C$_8$ configurations, the value of $E_a$
is $4\div 12$ times lower than that for $S=0$. Note that the dependence of
$E_a$ on $S$ in the range $S=1\div 7$ is nonmonotonous and has local maxima
at $S=4$ and $S=7$.

In attempts to get a physical insight into the nonmonotonous dependence of
$E_a$ versus $S$, we have calculated the difference
$\Delta=E_{LUMO}-E_{HOMO}$ between the energies of the lowest unoccupied and
the highest occupied molecular orbitals as a function of $S$. The results are
shown in Fig.5. One can see that the curve $\Delta (S)$ has maxima at $S=$ 0,
4, and 7, in accordance with higher values of $E_a$ for those values of $S$,
see Fig.4.

Now we turn to small ensembles composed of spin-polarized prismanes C$_8$. We
have studied various arrangements of two prismanes C$_8$ with respect to each
other and found four different metastable configurations (C$_8$)$_2$. They
are shown in Fig.6. In configuration A, the intercluster bonding occurs via
"top atoms", so that there is a single inter-cluster bond. In configurations
B and C, there are two inter-cluster bonds between "edge atoms", while in
configuration D, there are four inter-cluster bonds (the rectangular facets
of the prismanes are parallel). Note that prismanes in the ensembles
(C$_8$)$_2$ preserve their
overall original shape, they do not merge into a new cluster C$_{16}$ and
survive as the "building blocks" of the ensembles.

For either of four configurations (C$_8$)$_2$ shown in Fig.6, we have
restricted our studies to the range $S=0\div 10$. We have found that the
metastable configuration A exists at all values of $S=0\div 10$, while the
configurations B, C, and D are metastable, respectively, at $S=$ 0 and 7; 6
and 7; 8 and 9 only. The binding energy $E_b$ of (C$_8$)$_2$ ensembles
monotonously decreases with $S$ from $E_b=5.2$ eV/atom at $S=0$ down to
$E_b=3.4$ eV/atom at $S=10$, see the middle curve in Fig.3. The value of
$E_b$ at a given $S$ is almost the same in different metastable
configurations (C$_8$)$_2$, if any.

Although at $S\neq 0$ the curve $E_b(S)$ for the ensembles (C$_8$)$_2$ lies
well above that for the prismane C$_8$, see Fig.3, a close inspection of
those curves shows that the binding energy $E_b(2S)$
of an ensemble (C$_8$)$_2$ composed
of two prismanes C$_8$ with the spin $S$ each is almost the same as the
binding energy $E_b(S)$ of an isolated prismane C$_8$ with the spin $S$,
being higher by just $(0.1\div 0.2)$ eV/atom for all values of $S$
considered. Hence, the energy $E_{acc}=E_b^{graphite}-E_b$ accumulated in two
metastable prismanes C$_8$ changes insignificantly upon formation of the
ensemble (C$_8$)$_2$. A small increase in $E_b$ by $(0.1\div 0.2)$ eV/atom is
due to appearance of new covalent bonds between two prismanes C$_8$ and, as a
consequence, the decrease of the total energy of the system. The bond energy
is about 1 eV/bond.

In contrast to the prismane C$_8$, the activation energy $E_a$ for
the decay of the ensemble (C$_8$)$_2$ in configuration A as a
function of $S$ has a broad maximum, $E_a=(0.35\pm 0.05)$ eV, at
$S=6\div 8$, while the value of $E_a$ for the non-polarized
ensemble (C$_8$)$_2$ with $S=0$ is very low, $E_a\approx 0.03$ eV,
see Fig.7. (We recall that in the prismane C$_8$, the maximum of
the $E_a(S)$ curve is at $S=0$). For other metastable (C$_8$)$_2$
configurations shown in Fig.6, the value of $E_a$ equals to
$(0.83\pm 0.12)$ eV and $(0.12\pm 0.03)$ eV for configuration B
with $S=$ 0 and 7 respectively; $(0.03\pm 0.005)$ eV and $(0.29\pm
0.03)$ eV for configuration C with $S=$ 6 and 7 respectively;
$(0.1\pm 0.03)$ eV and $(0.07\pm 0.01)$ eV for configuration D with
$S=$ 8 and 9 respectively. It is worth noting that several
spin-polarized configurations (C$_8$)$_2$ with specific values of
$S$ appear to be more stable than constituting spin-polarized
prismanes C$_8$, c.f. Figs. 4 and 7.

We have also studied the metastability of quasi-one-dimensional
spin-polarized ensembles (C$_8$)$_3$ formed upon attachment of a
third prismane C$_8$ to the ensemble (C$_8$)$_2$ in configuration
A, see Fig.8. We have restricted ourselves to the range $S=0\div
16$. We have found that the metastable configuration (C$_8$)$_3$
exists at all values of $S$ in this range. The binding energy
$E_b$ of the (C$_8$)$_3$ ensemble monotonously decreases with $S$
from $E_b=5.3$ eV/atom at $S=0$ down to $E_b=3.3$ eV/atom at
$S=16$, see the upper curve in Fig.3. From Fig.3 one can find that
the binding energy $E_b(3S)$ of an ensemble (C$_8$)$_3$ composed of
three prismanes C$_8$ with the spin $S$ each exceeds the binding
energy $E_b(S)$ of an isolated prismane C$_8$ with the spin $S$ by
$\approx 0.2$ eV/atom for all values of $S$ in the range $S=1\div
5$. Hence, an increase in the number $m$ of prismanes in a
metastable spin-polarized ensemble (C$_8$)$_m$ leaves the energy
$E_{acc}=E_b^{graphite}-E_b$ accumulated in the ensemble almost
unchanged.

The shape of the curve $E_a(S)$ for the ensemble (C$_8$)$_3$ is similar to
that for the ensemble (C$_8$)$_2$, see Fig.7. Note, however, that the maximum
of $E_a(S)$ for the ensemble (C$_8$)$_3$ is shifted to greater values of
$S=$ 9, 10 and is higher by $\approx 0.1$ eV than in the case of the
ensemble (C$_8$)$_2$. So, the stability (and hence, the lifetime) of the
system increases with the number of prismanes in the ensemble.

Fig.9 shows the activation energy $E_a$ versus the difference
$\Delta=E_{LUMO}-E_{HOMO}$ between the energies of the lowest unoccupied and
the highest occupied molecular orbitals for all metastable configurations
considered in this work. One can see the overall increase of $E_a$ with
$\Delta$. Such a correlation between $E_a$ and $\Delta$ indicates that the
nonmonotonous dependence of $E_a$ on $S$ stems from the nonmonotonous
dependence of $\Delta$ on $S$.

\vskip 6mm

{\bf 4. Conclusions}

\vskip 2mm

We have numerically examined a possibility of existence of a spin-polarized
cage-like metastable carbon cluster prismane C$_8$ and small ensembles
composed of those clusters. We have shown by molecular dynamics simulations
that the activation energy $E_a$ for the decay of a metastable configuration
is a nonmonotonous function of the total spin $S$ of the system and has
maxima at certain values of $S$. The most stable structures are characterized
by the large energy gap between the lowest unoccupied and the highest
occupied molecular orbitals. The maximum value of $E_a$ increases
with the number of prismanes in the ensemble, pointing to a possibility of
existence of "cluster matter" composed of spin-polarized clusters C$_8$. The
metastable ensembles of prismanes C$_8$ accumulate a great amount of energy
(several eV/atom) that may be released upon fusion of a large number of
prismanes into the bulk specimen. Hence, such ensembles can be considered as
candidates to the high energy density materials (HEDM). The role of spin-flip
processes for the stability of spin-polarized cluster configurations needs
further consideration.

\vskip 6mm

{\bf Acknowledgments}

\vskip 2mm

We are grateful to V. F. Elesin and N. N. Degtyarenko for valuable
discussions and A. I. Podlivaev for the help in numerical calculations.
The work was supported by the CRDF (project "Basic studies of matter in
extreme states") and by the Russian Federal Program "Integration"
(project No A0133).


\newpage
\centerline{\bf Figure captions}
\vskip 2mm

Fig.1. Metastable prismane C$_8$. For $S$ = 0, 4, and 7 the prismane
structure is characterized by three different bond lengths $a$=AE=BF=CG,
$b$=AD=BD=CD=EH=FH=GH, and $c$=AB=AC=BC=EF=EG=FG (symmetrical configuration).
For other values of $S$, the prismane is slightly distorted, see text and
Fig.2 for details. $a=1.28$ {\AA}, $b=1.47$ {\AA}, $c=2.31$ {\AA} for $S=0$.

Fig.2. Bond lengths $a$ (circles), $b$ (triangles), and $c$ (squares) of
metastable prismane configurations versus the total spin $S$, see Fig.1 for
the definition of $a, b, c$. For the distorted configurations with
$S = 1, 2, 3, 5, 6$ there are more than three different bond lengths, i.e.,
some of equalities AE=BF=CG, AD=BD=CD=EH=FH=GH, AB=AC=BC=EF=EG=FG
are violated.

Fig.3. Binding energy $E_b$ of the metastable prismane C$_8$ (closed
circles), see Fig.1, the metastable ensembles (C$_8$)$_2$ in configurations A
(closed triangles), B (open circles), C (open squares), D (open triangles),
see Fig.6, and the metastable ensemble (C$_8$)$_3$ (closed squares),
see Fig.8 versus the total spin $S$ of the system. Lines are the guides to
the eye.

Fig.4. Activation energy $E_a$ for the decay of the metastable prismane
C$_8$ versus the total spin $S$. Line is the guide to the eye.

Fig.5. The difference $\Delta=E_{LUMO}-E_{HOMO}$ between the energies of the
lowest unoccupied and the highest occupied molecular orbitals of the
metastable prismane C$_8$ versus the total spin $S$. Line is the guide to the
eye.

Fig.6. Four different configurations (A, B, C, D) of metastable ensembles
(C$_8$)$_2$.

Fig.7. Activation energy $E_a$ for the decay of the metastable
ensemble (C$_8$)$_2$ in configuration A (circles), see Fig.6a, and
quasi-one-dimensional ensemble (C$_8$)$_3$ (triangles), see Fig.8, versus the
total spin $S$ of the system. Lines are the guides to the eye. Two values of
$E_a$ of the ensemble (C$_8$)$_2$ at $S=1$ correspond to two basically
different paths of the ensemble decay.

Fig.8. Quasi-one-dimensional metastable ensemble (C$_8$)$_3$.

Fig.9. Activation energy $E_a$ for the decay of metastable spin-polarized
configurations versus the difference $\Delta=E_{LUMO}-E_{HOMO}$ between the
energies of the lowest unoccupied and the highest occupied molecular
orbitals for the metastable prismane C$_8$ (closed circles), the metastable
ensembles (C$_8$)$_2$ in configurations A (closed triangles),
B (open circles), C (open squares), D (open triangles), and the metastable
ensemble (C$_8$)$_3$ (closed squares).

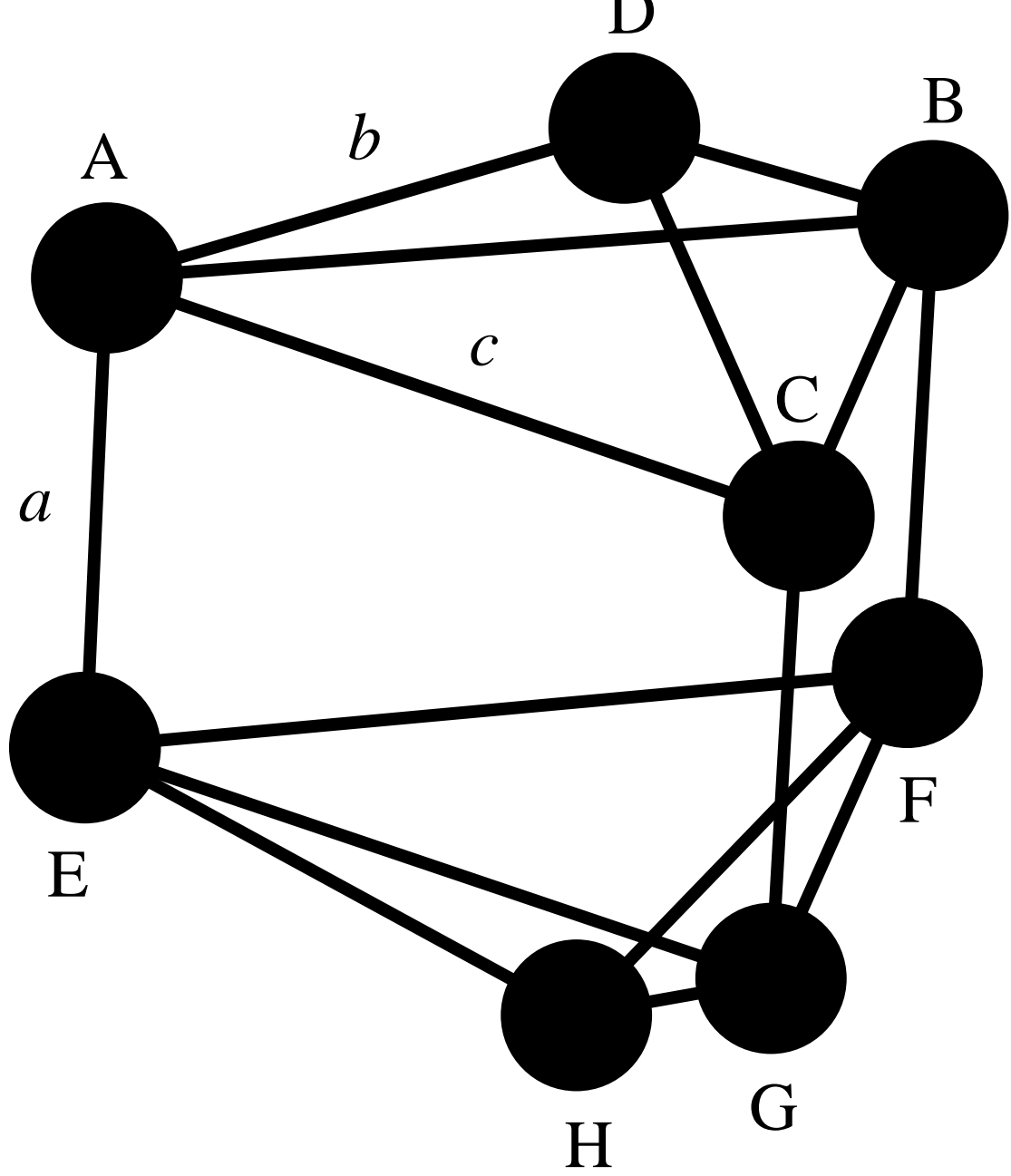
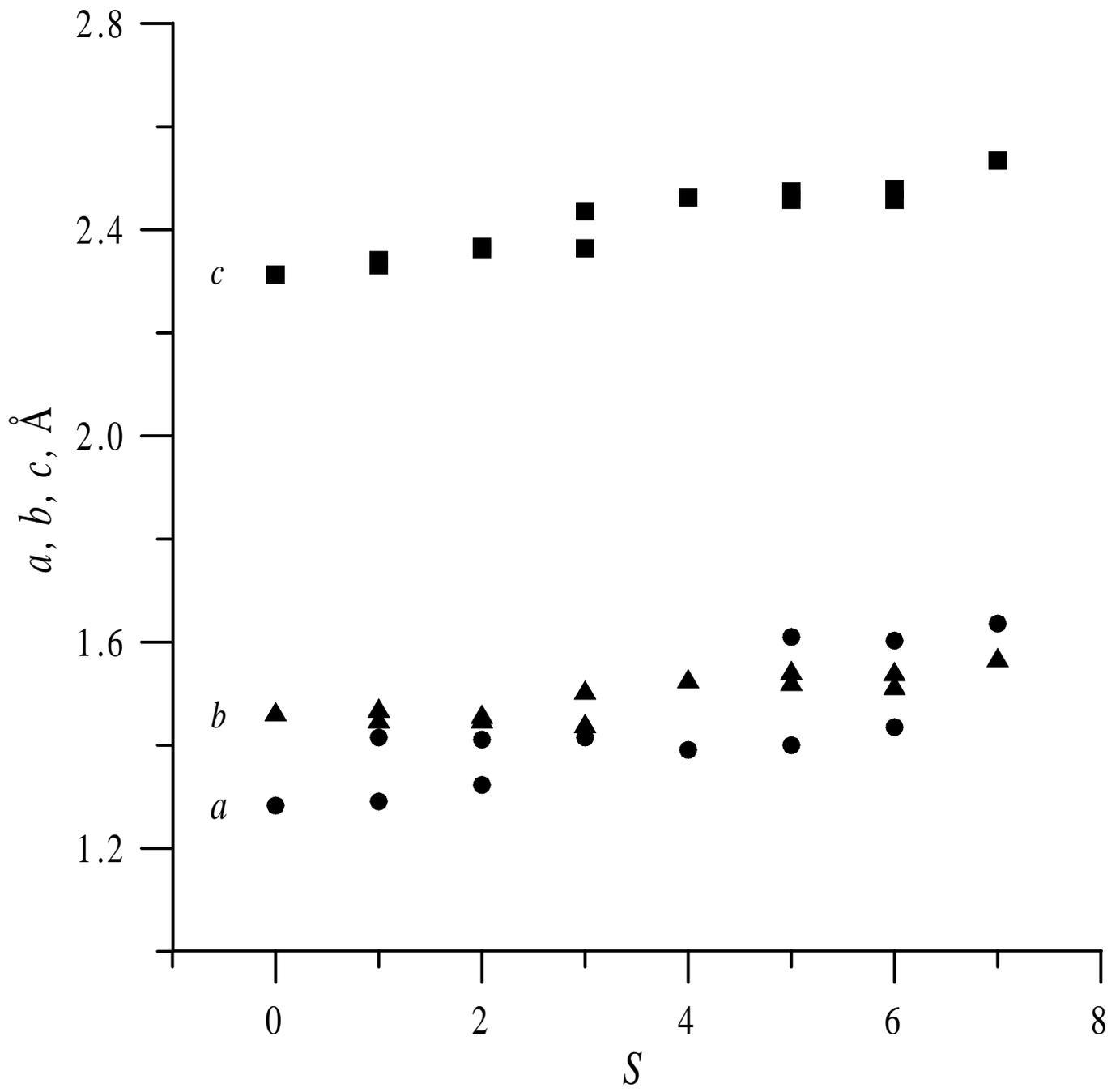
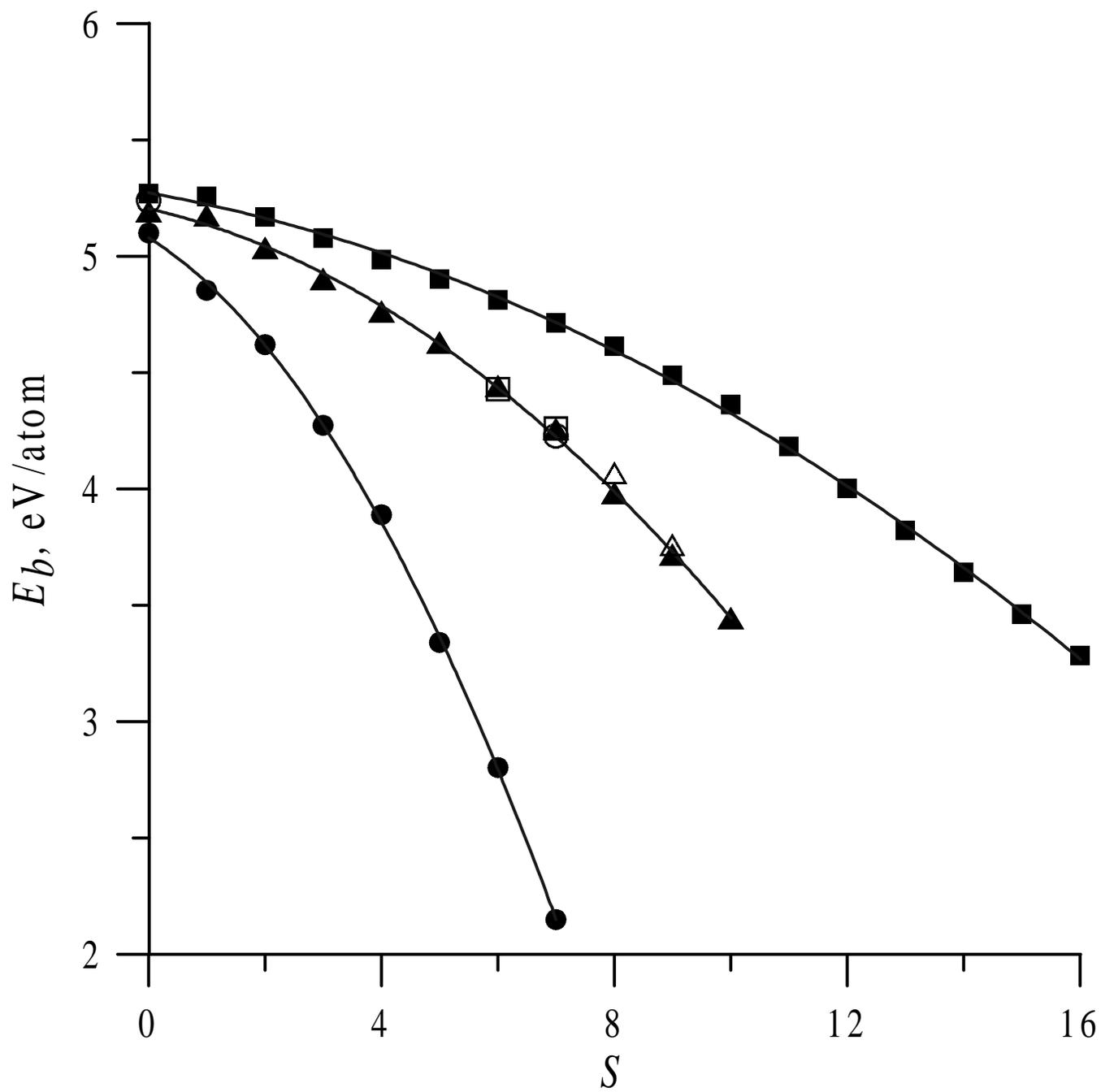
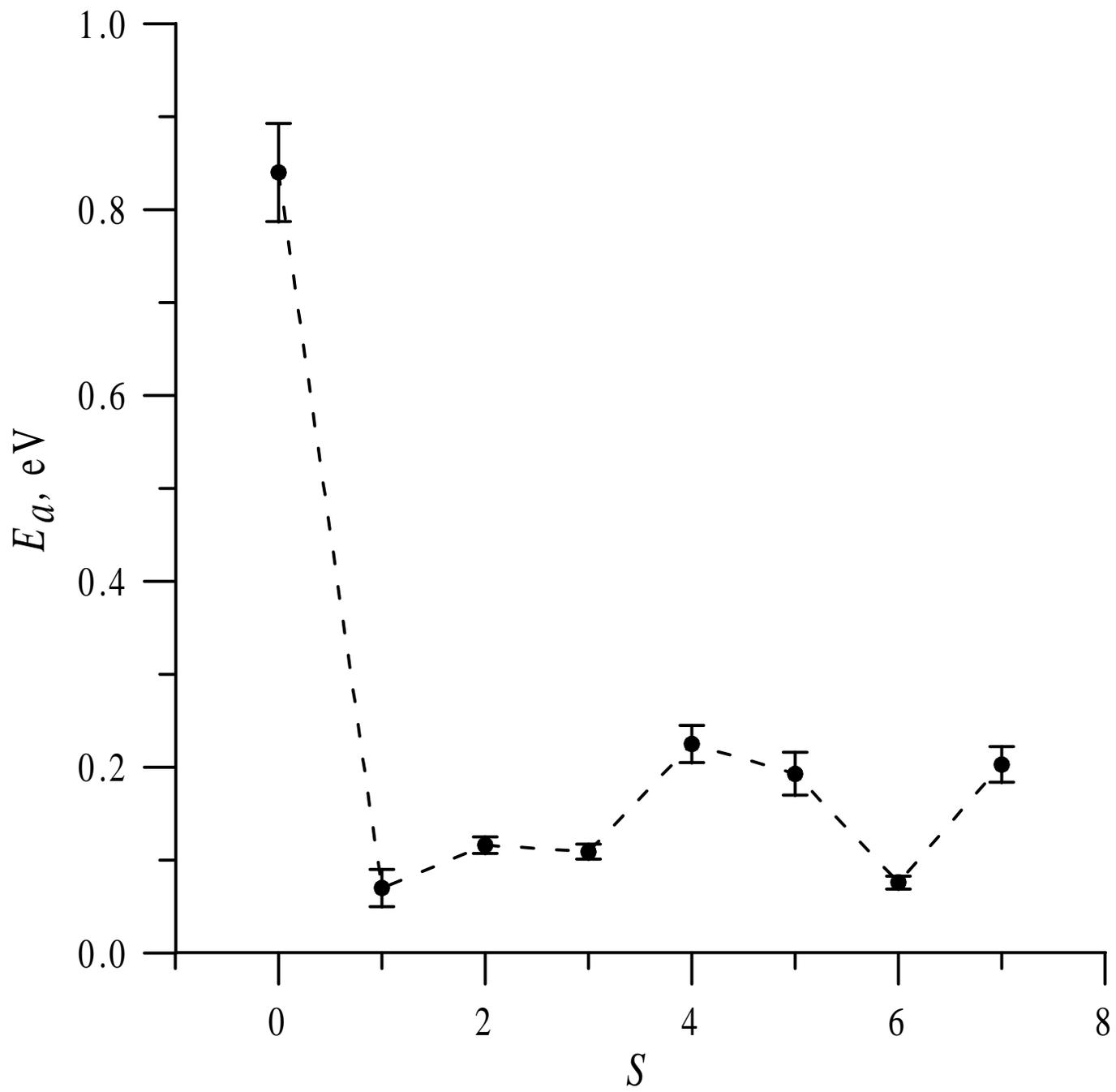
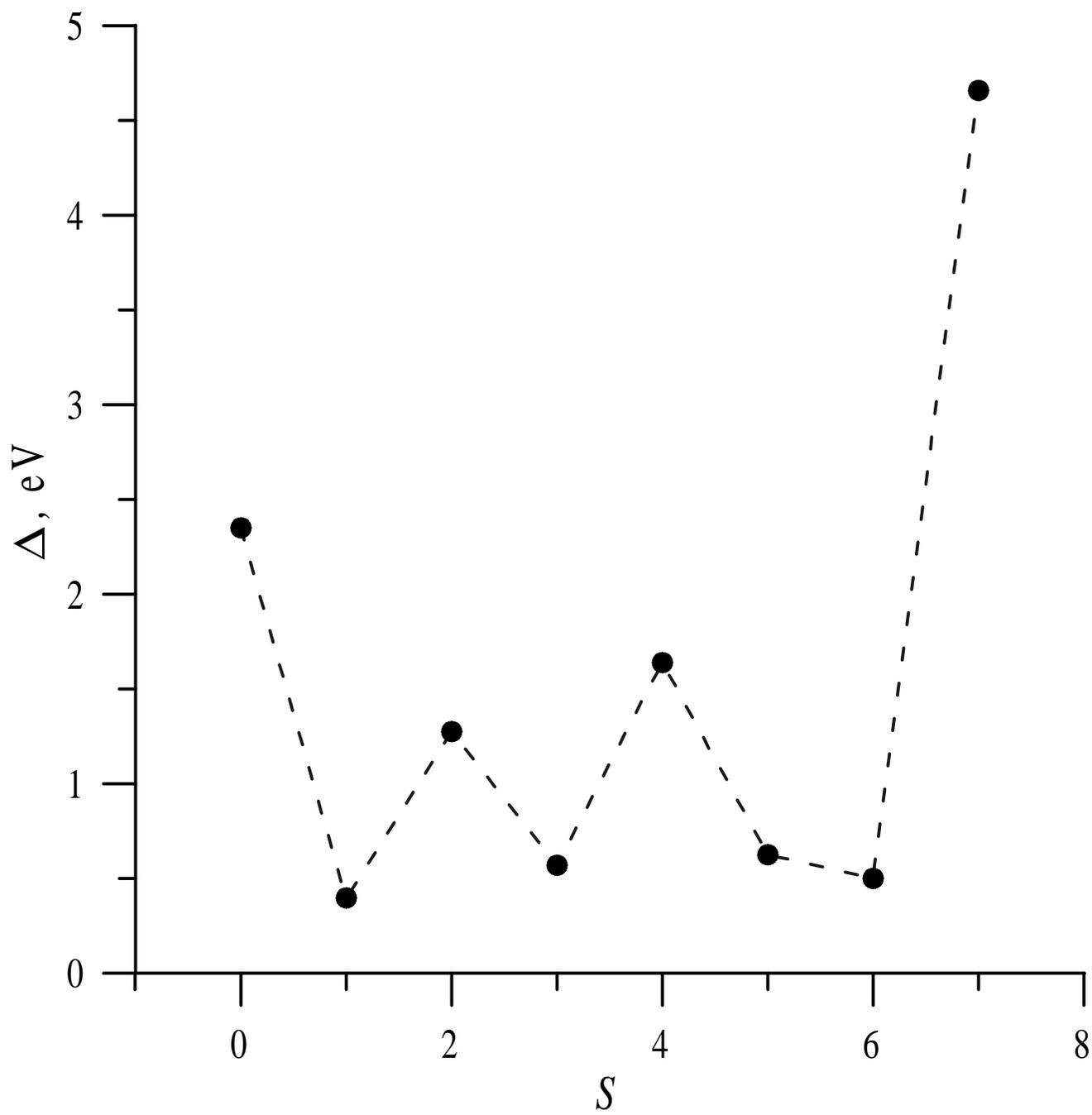
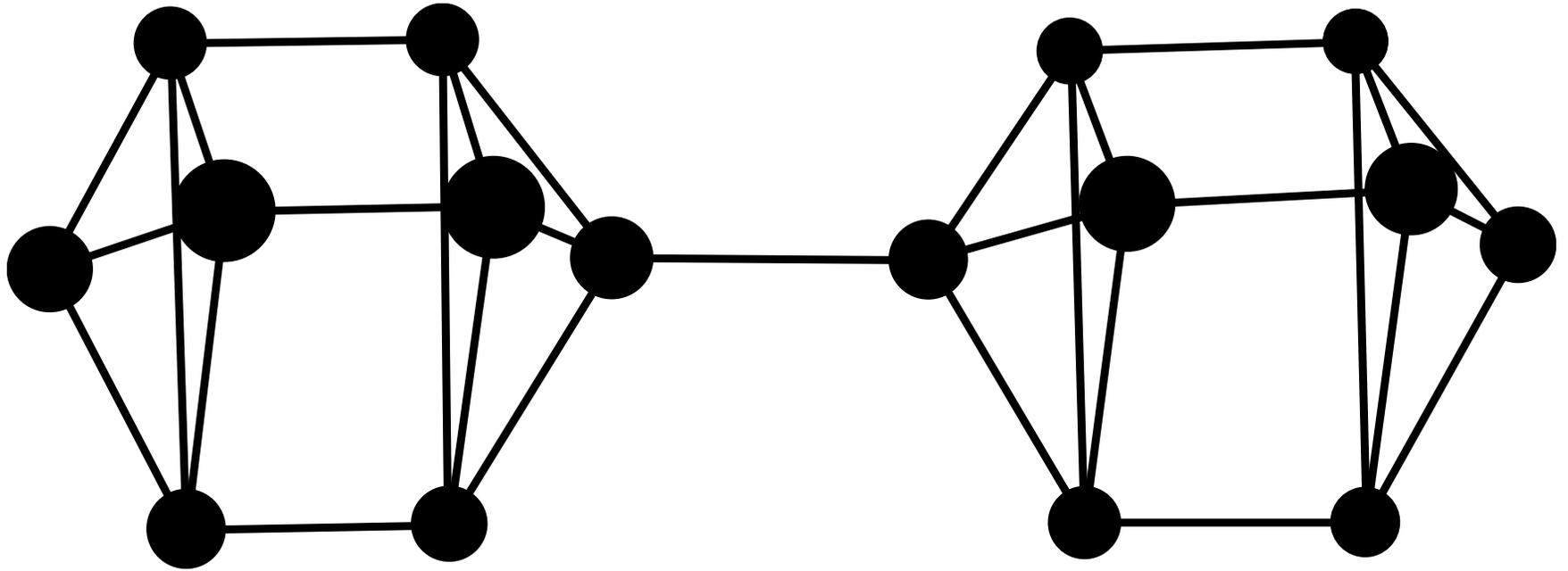
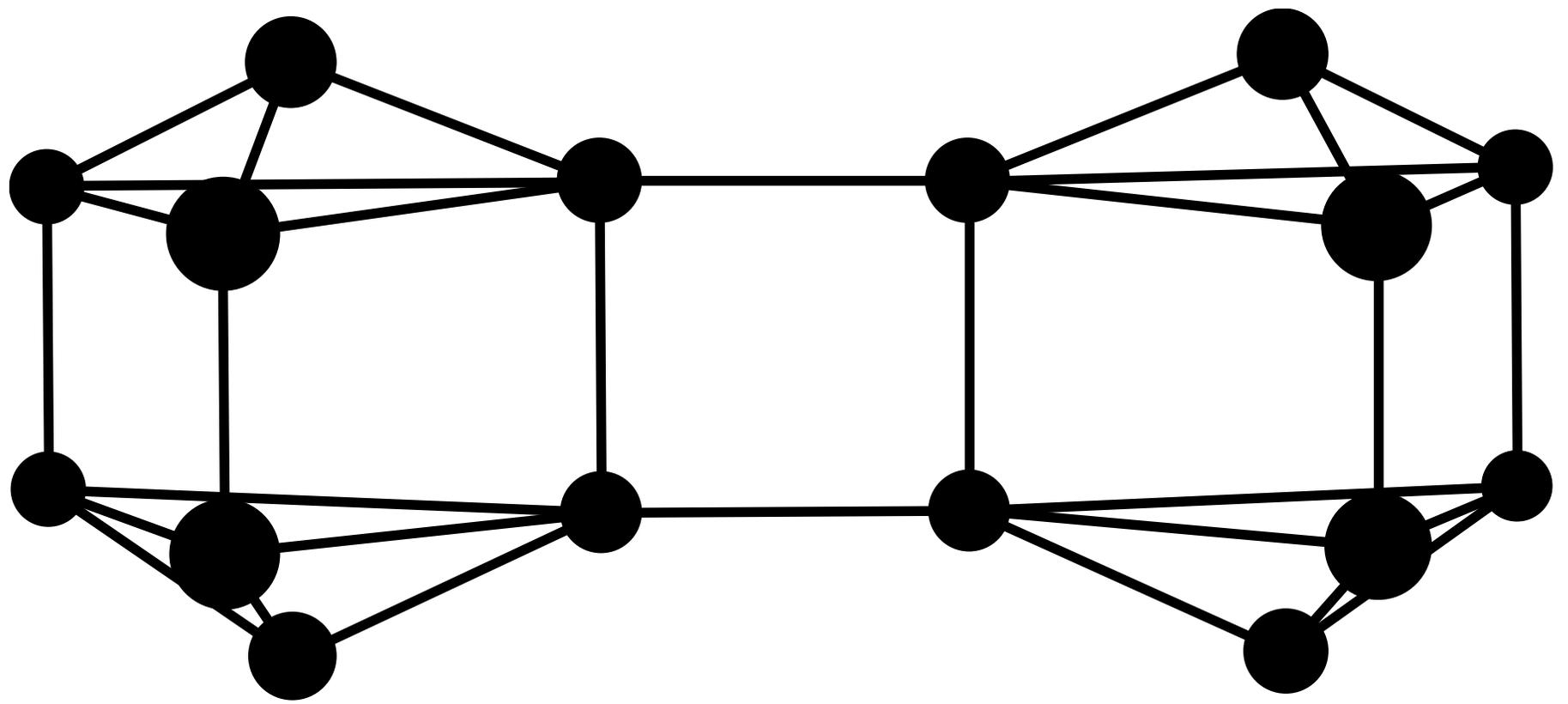
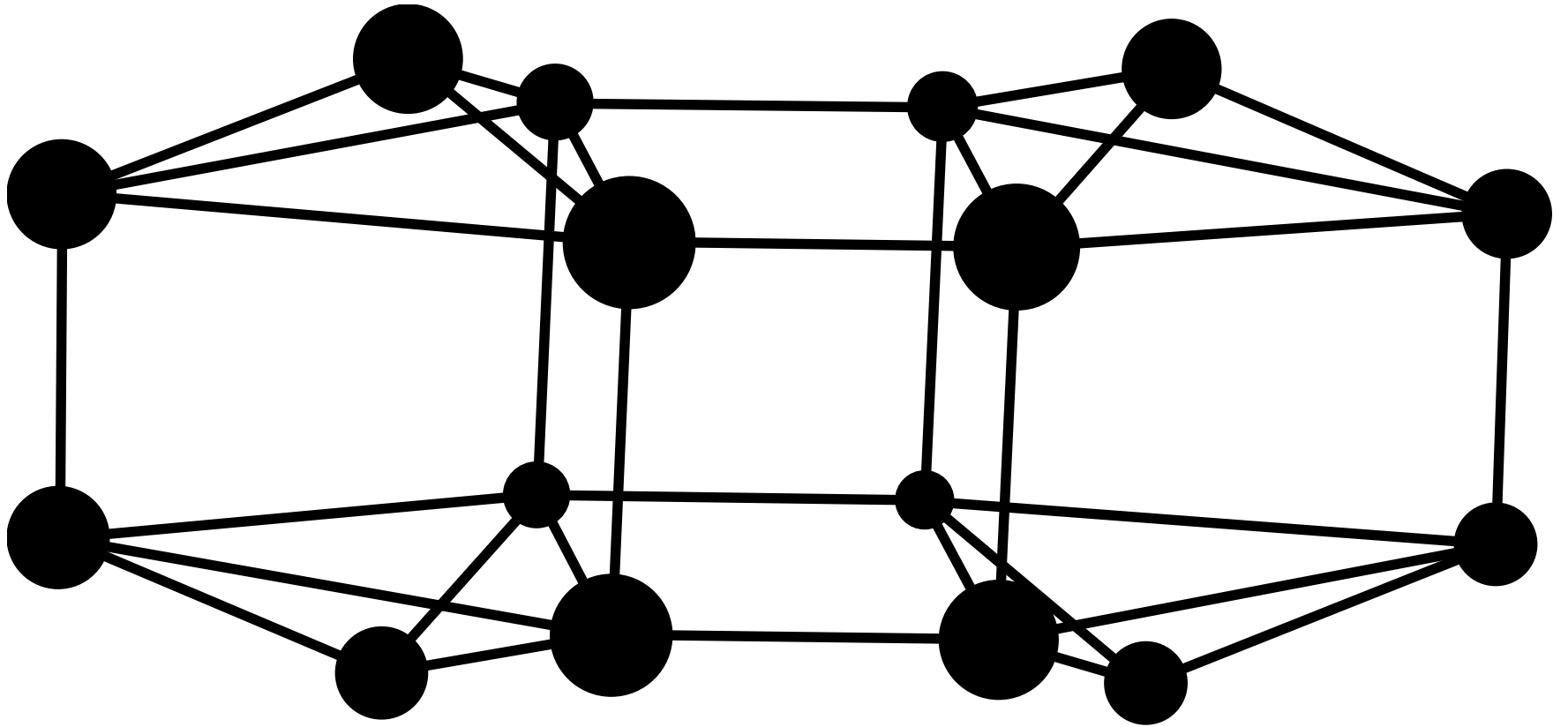
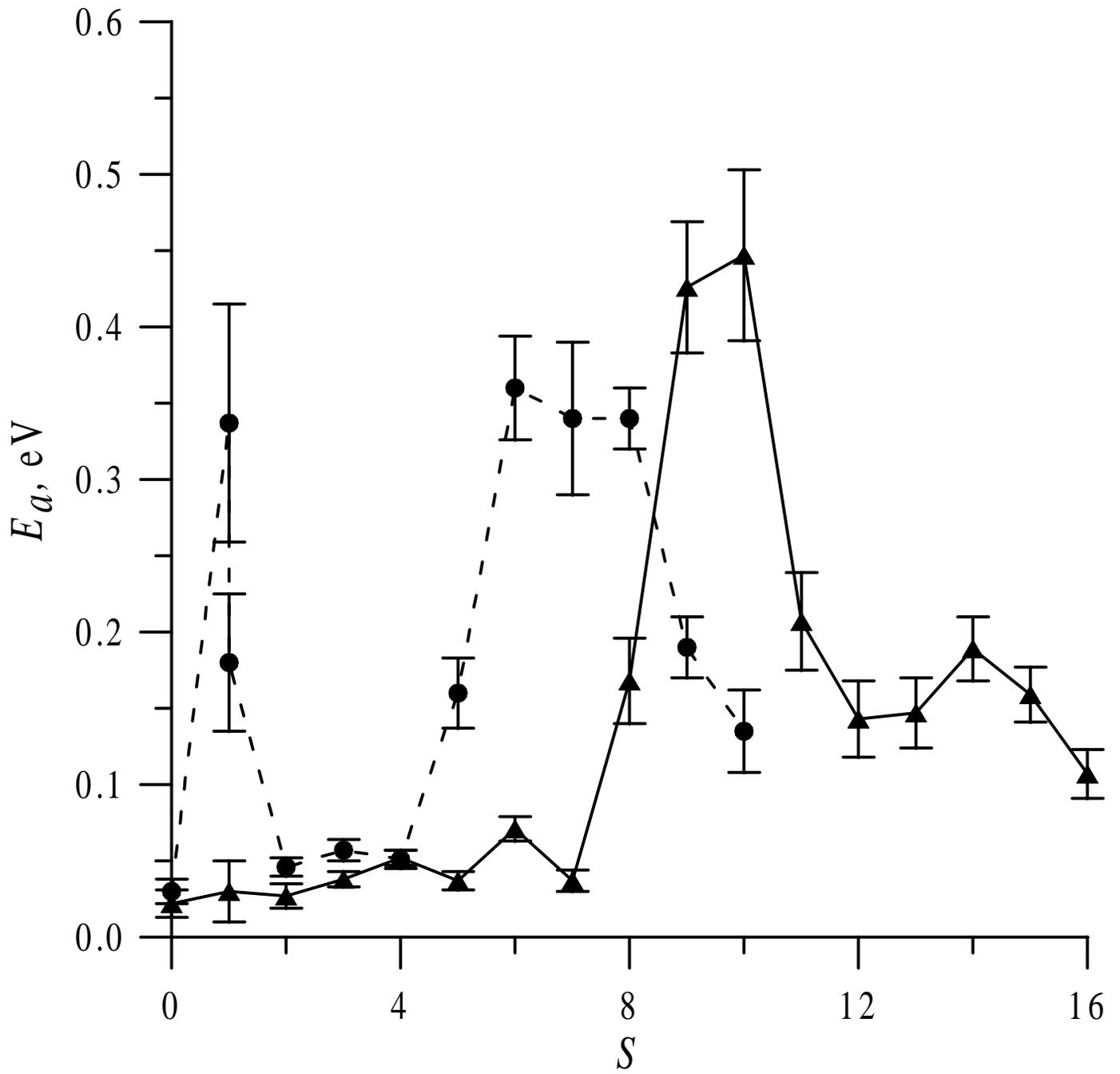
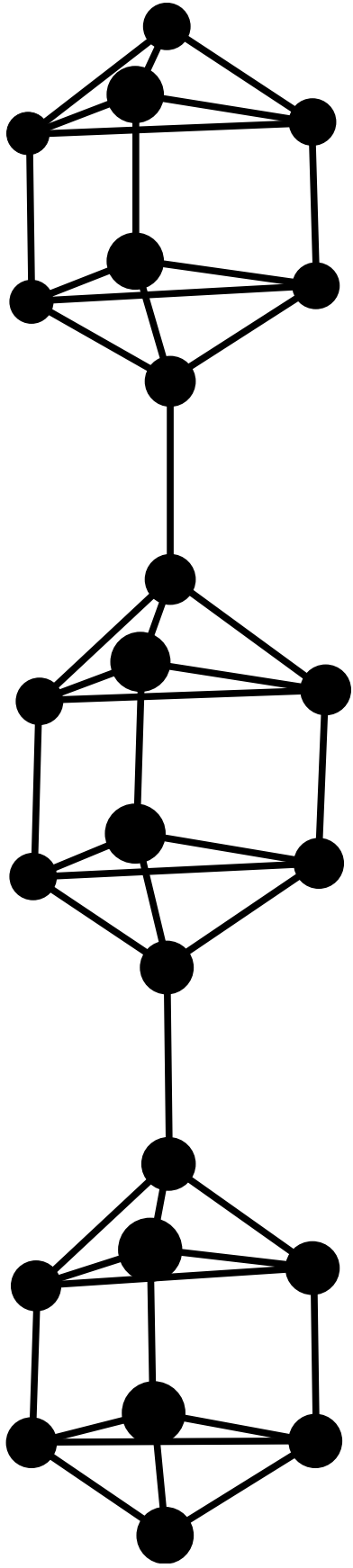
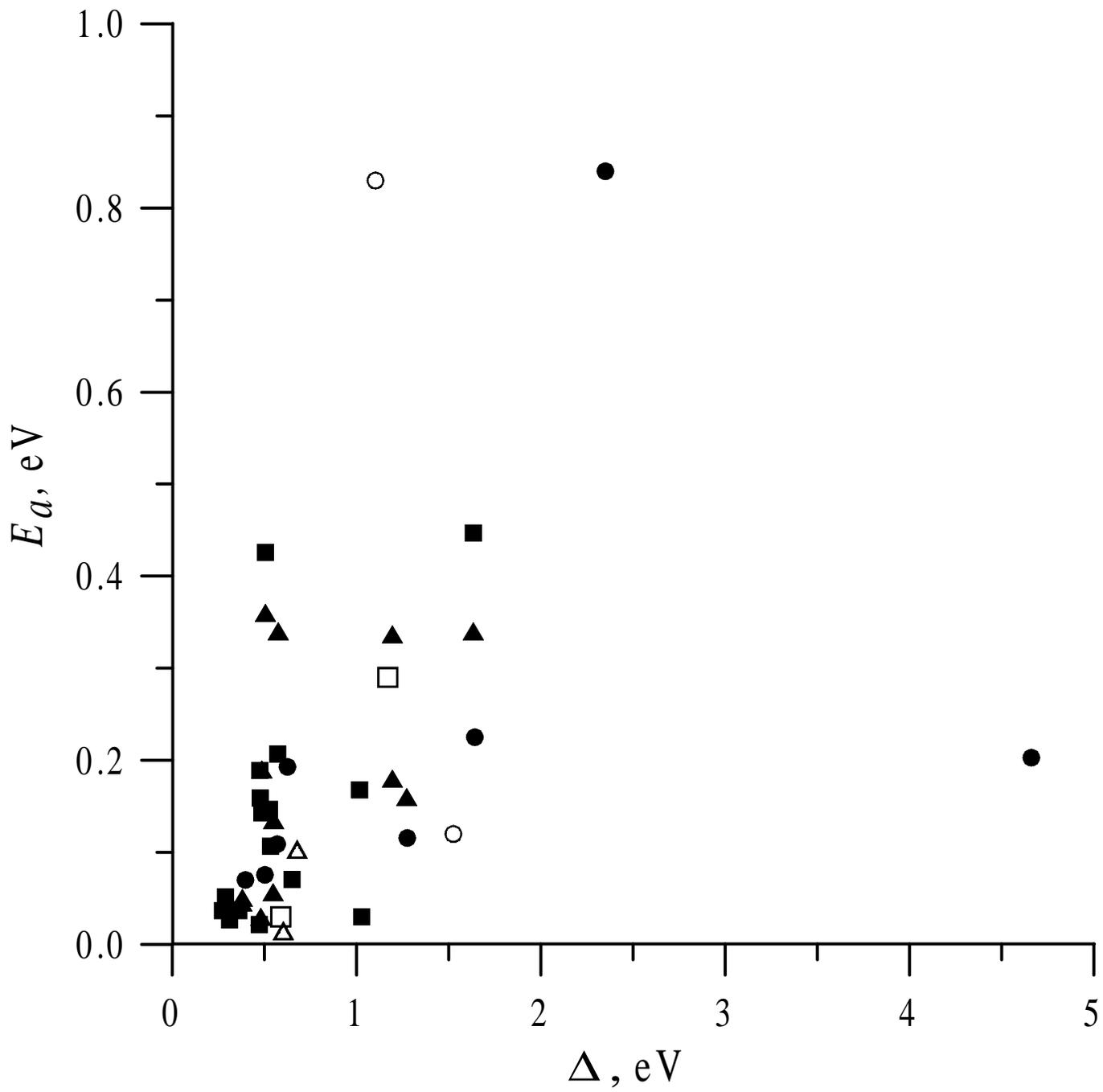
\end{document}